# Dark Metastable Conduction Channels near a Metal-Insulator Transition

T. R. Devidas[1], Dror Yahav[4], Jonathan T. Reichanadter[2,3], Matan Sterenberg[4], Shannon C. Haley[2,3], Joel E. Moore[2,3], Jeffrey B. Neaton[2,3], James G. Analytis[2,3,5,*], Beena Kalisky[1,*], Eran Maniv[4,*]

[1]Department of Physics and Institute of Nanotechnology and Advanced Materials, Bar-Ilan University, Ramat Gan 5290002, Israel
[2]Department of Physics, University of California, Berkeley, California 94720, USA
[3]Materials Science Division, Lawrence Berkeley National Laboratory, Berkeley, California 94720, USA
[4]Department of Physics, Ben-Gurion University of the Negev, Beer-Sheva 84105, Israel
[5]CIFAR Quantum Materials, CIFAR, Toronto, Canada

* Corresponding authors: Email: analytis@berkeley.edu, beena@biu.ac.il, eranmaniv@bgu.ac.il

**Materials that transition between metal and insulator, the two main states that distinguish all solids, are fascinating because they underlie many mysteries at the frontier of solid state physics. In 1T-TaS$_2$, the metal-insulator transition is linked to a metastable hidden state arising within a chiral charge density wave (CDW) whose basic nature remains an open question. In this work, we show that pulses of current through these materials create current-carrying filamentary channels that distinguish the 'metallic' hidden state and 'insulating' CDW states. These channels have remained dark to previous measurements, and yet are directly linked to the properties of the hidden state. We leverage the metastability of these conduction channels to demonstrate electrical control of their creation, erasure and location. Our findings show that physical elements, such as boundaries and interfaces, play a key role in the properties of the hidden state characterizing the metal-insulator transition. We suggest new possibilities for in-situ electrical design of synaptic components with possible applications to neuromorphic computing**.

Materials at the boundary of a metal-insulator transition are of significant interest in the field of condensed matter physics. They include high temperature superconductors and spin liquids, fractional Chern insulators and topological superconductors. The challenge of understanding these materials is often that they are found at intermediate energy scales, where different channels of interaction effectively compete to set the quantum state of the system. This complexity undermines the central assumptions we make to understand materials and particularly metals: the existence of well-defined, charge carrying "quasiparticles" which arise from the ability of electrons to screen and behave as "dressed" (as opposed to bare) electrons, and ultimately leading to simple linear responses like the familiar Ohm's law connecting current to applied electric fields. In the exotic materials at the frontier of research, it is these assumptions that break down, and in this situation we can do little but resort to giving such materials labels that distinguish them, like "strange metals", "pseudogaps" or "hidden states". In truth, the main virtue of such labels is largely that they help articulate how the complexity of the problem has evaded our best attempts at explanation.

One such material for which “hidden states” appear near a metal-insulator transition is the transition-metal dichalcogenide (TMD) 1T-$TaS_2$ (*1*). This layered metal orders at 550 K into an Incommensurate Charge Density Wave (I-CDW) phase, followed by a nearly commensurate CDW (N-CDW) ordering at ~350 K (*2*, *3*). It then undergoes a metal-insulator transition below ~180 K that induces the formation of a commensurate CDW (C-CDW) lattice composed of $\sqrt{13}\times\sqrt{13}$ Star-of-David (SD) hexagram atomic distortions, as illustrated in the inset of Fig. 1(A). Of these phases, only the C-CDW is an insulator, and all others are metallic. This state has been proposed to result from a Mott transition, where Coulomb interactions drive electron localization (*4–6*). The C-CDW state of 1T-$TaS_2$ can be driven into a metallic state using large electric fields or laser pulsing (*1*, *7*, *8*). This state is neither a typical metal nor a typical insulator; its resistance is Ohmic (like a metal) but decreases with increasing temperature (like an insulator) and so has been described as hosting an underlying ‘hidden state’ (*1*, *8*). It is not clear whether the hidden state is simply a manifestation of the N-CDW; notably, if flakes of the material are super-cooled from the N-CDW state they remain metallic, an effect that could be mimicked by laser or current-induced Joule heating (*9*). On the other hand, scanning tunneling microscopy (STM) studies of electrically pulsed 1T-$TaS_2$ nano-flakes have suggested that the conducting and insulating states arise from a quasi-continuum of ‘metastable states’ (*10*) related to the density and topology of the domain structure in the underlying CDW (*11–13*). This hidden state disappears at around $T_H$ ~ 70 K, a temperature scale that is mysterious because it does not correspond to the formation of any of the CDW phases described above. The nature of the hidden state and its energy scale, is the central question we address in this study.

The hidden state can be accessed through laser pulses or the application of large electric fields. In this study we use large direct current pulses to drive the insulator-metal transition into the hidden state. We fabricate devices using Focused Ion Beam (FIB) microstructuring, creating cross-bar structures as shown in the inset of Fig. 1A. The current-driven metal-insulator is manifested as a dramatic decrease in the resistivity as compared to the cooldown curve, as shown in Fig. 1A, which occurs above a certain threshold current density of approximately $I_{write} \sim 6 \times 10^4$ A/cm$^2$ (for device 1). This basic behavior is highly reproducible between devices, as shown for a separate device (device 2) in Fig. 1B. Up until a threshold density $I_{write} \sim 10^5$ A/cm$^2$ (in device 2) is reached, the system remains in the high resistance (insulating) state. The low resistance state has all the hallmarks of the hidden state; a significantly lower resistance which disappears at $T_H$ ~ 70 K, the characteristic hidden state transition temperature $T_H$. Interestingly, an intermediate current of $I_{erase} \sim 5 \times 10^4$ A/cm$^2$ returns the system to its high resistance state, so that the device can be electrically switched between low and high resistive states reproducibly. This behavior is reminiscent of previous experiments using voltage pulses on flakes of 1T-$TaS_2$, where high and intermediate voltage pulses manifest similar behavior (*14*, *15*) and ascribed to some mysterious interplay of non-equilibrium C-CDW dynamics and Joule heating.

To gain insight into this interplay, we employ scanning-SQUID magnetometry and image the current directly (see Methods for details). Scanning-SQUID magnetometry is a powerful probe of local magnetic signals at intermediate scales, bridging bulk and atomic-scale experimental techniques. In Fig. 1C we show the current paths derived from images of the flux generated by the alternating current ($I_{read}$) flow (local Biot-Savart fields, see supplementary Fig. S3). On the scale of the resolution of the experiment (~ 2 μm), the current appears to flow homogeneously through the bulk of the sample (supplementary Fig. S4). After the application of “pulse” current ($I_{write}$) beyond the threshold, which brings the system to the hidden state, the current path is completely localized to the physical boundary, as shown in Fig. 1D. These remain indefinitely (supplementary Fig. S5) unless the system is warmed above $T_H$ (supplementary Fig. S6 C-D), confirming the connection to the hidden state. The current in the hidden state is mediated by non-volatile, confined conduction paths spanning the length of the device.

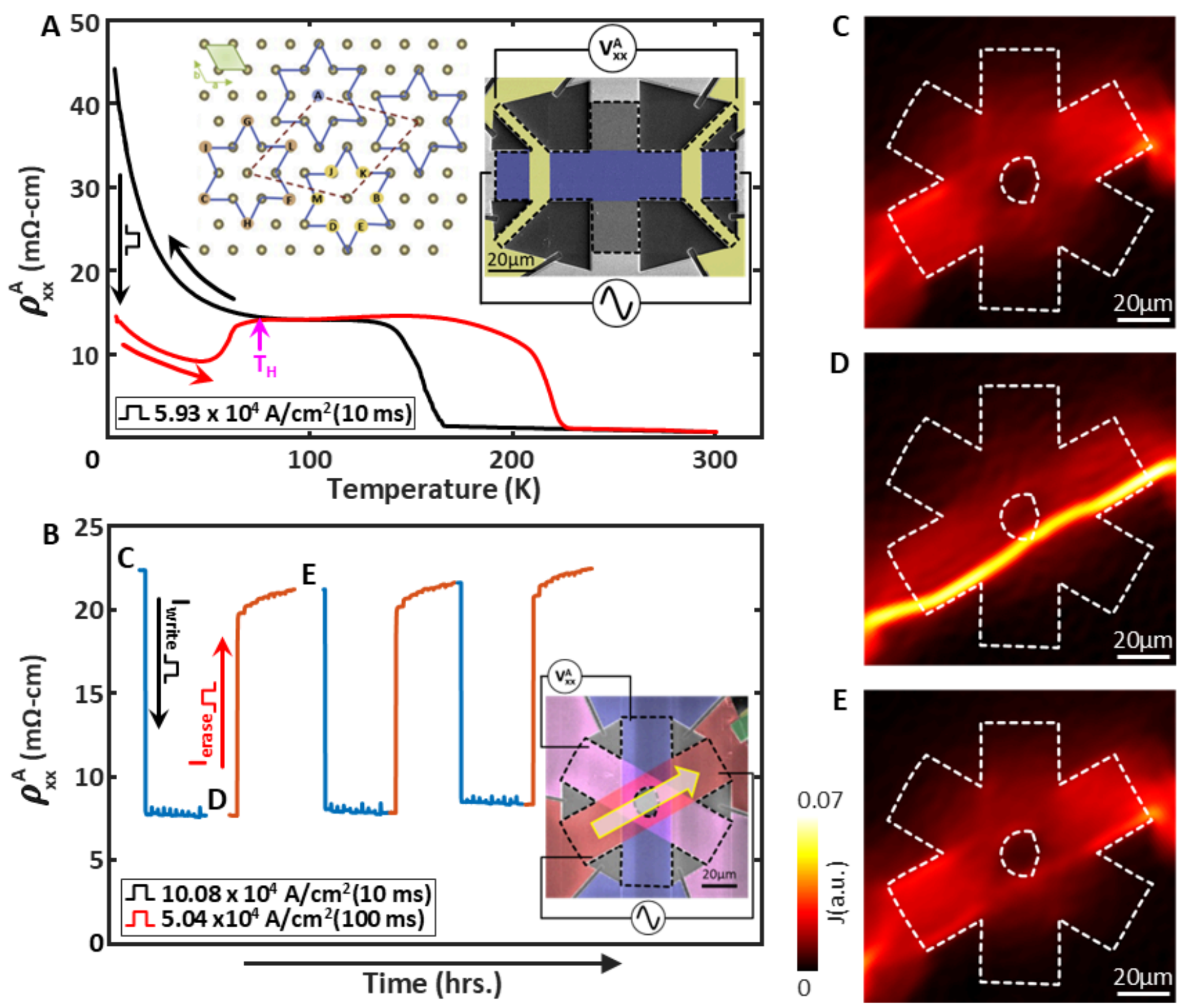

**Fig. 1. Writing and erasing DMCs in 1T-$TaS_2$.** (**A**) Temperature dependent a.c. transport data in 1T-$TaS_2$ (device 1). The black curve is cooldown data from room temperature to 4.2 K. The N-CDW to C-CDW transition is observed. The black square pulse indicates the application of a direct current pulse ($I_{write}$ = 5.93 x $10^4$ A/cm$^2$, 10ms) at 4.2 K, 10ms being the pulse-width. The black arrow indicates resistivity decrease after application of $I_{write}$ pulse. The system switches into a "hidden state". The red curve is the warmup data from 4.2 K to room temperature after pulsing. Left inset: Unit cell and star-of-David (SD) charge density wave (CDW) structure in 1T-$TaS_2$. Only Tantalum ions are depicted for visual clarity; right inset: SEM image of device 1 with the 4 probe a.c. resistivity measurement configuration, the source-drain channel being indicated by the false color. (**B**) Alternate switching in transport between the C-CDW state and the hidden state in 1T-$TaS_2$ (device 2) by application of d.c. pulses. $I_{write}$ (10.08 x $10^4$ A/cm$^2$, 10ms) switches the device from as-cooled C-CDW state with high resistivity into the hidden state with low resistivity. $I_{erase}$ (5.04 x $10^4$ A/cm$^2$, 100ms), switches the resistivity from low to high value. Inset: SEM of device 2 with false colors indicating the different pulse leads. The source/drain leads for the 4-probe a.c. resistivity and the d.c. pulse leads are along the red colored channel. The white arrow with yellow outline indicates the polarity of the d.c. pulse. (**C-E**) Reconstructed current density maps (from scanning SQUID imaging, see supplementary Fig. S3) in device 2 in the different states by flowing an alternating current ($I_{read}$ = 12.61 A/cm$^2$). (C) is the as-cooled C-CDW map with current flowing along the entire channel, (D) is the map of the hidden state after $I_{write}$ pulsing. The current now flows in a confined channel between the leads. (E) is the map of the recovered

high resistivity C-CDW state. The confined channel is "erased" by the $I_{erase}$ pulse and current flows along the entire channel.

The resistivity in the low resistance state shown in Fig. 1B therefore greatly overestimates the true resistance of the conduction channel by at least a factor of 5; these are conducting highways compared to the residual conduction of the C-CDW state. We cycle the pulse sequence and when crossing an intermediate threshold of $I_{write}$ ~ 5x10$^4$ A/cm$^2$, we find the system returns to the high resistance state. Scanning-SQUID current maps show the current paths are indistinguishable from the first cooldown, as shown in Fig. 1E. The device can be reproducibly switched in this way between the metastable hidden and insulating ground states. The magnitude of this change in resistivity varies between devices with different geometries, ranging between a factor of 2-4. These conduction channels are only visible at these mesoscopic lengthscales, and dark to bulk or atomic-resolution probes, motivating us to describe them as dark metastable conduction channels (DMCs).

To distinguish the physics of the DMCs from the supercooled N-CDW state, we cool our devices at different cooling rates and study their response to current pulses. By controlling the cooling rate, the device either passed through (Fig. 2A, red) or suppressed (blue) the N-CDW to C-CDW phase transition, allowing us to test its response in both phases to $I_{write}$ current pulses. The supercooled N-CDW state has a significantly lower resistivity than the slow-cooled C-CDW state (Fig. 2A). The current profile in both the supercooled N-CDW (Fig. 2B) and slow-cooled C-CDW states (Fig. 2C) are homogenous. On application of $I_{write}$ pulses, the resistance of the supercooled N-CDW state (Fig. 2D) does not change, and the current profile remains the same (Fig. 2E). The supercooled state does not host the DMCs that characterize the hidden state that are observed in the slow-cooled C-CDW phase (Fig. 2F). This establishes that the underlying nature of the hidden state is distinct from the supercooled N-CDW. The DMCs can be borne only from the C-CDW ground state.

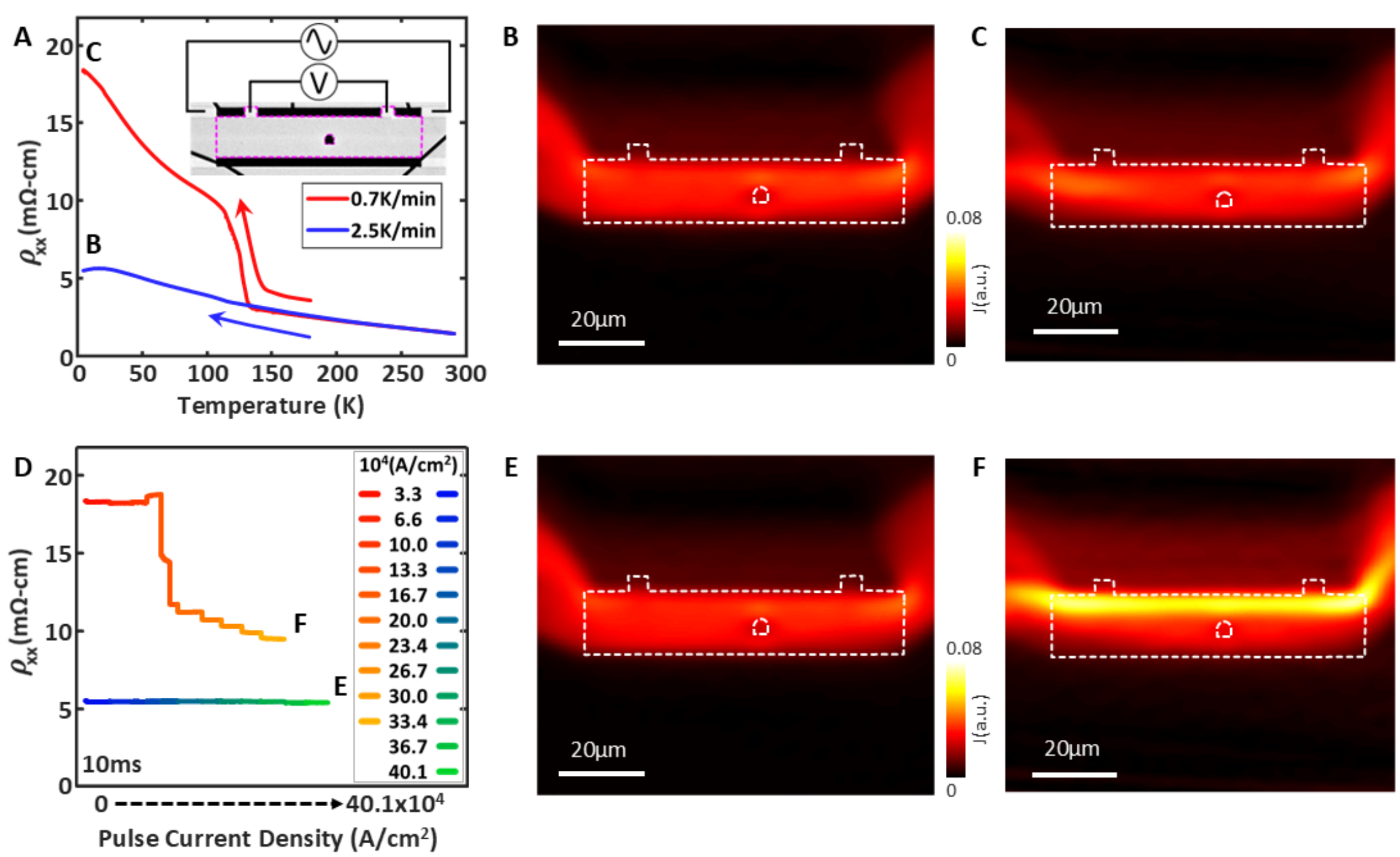

**Fig. 2. DMCs originate from the C-CDW phase.** (**A**) Temperature dependent a.c. transport data in 1T-$TaS_2$ (device 3-a) with two different cooling rates. When cooled from room temperature to 4.2 K at 0.7 K/min. (red), the device undergoes the N-CDW to C-CDW phase transition. When cooled from room temperature to 4.2 K at 2.5 K/min. (blue), the system is supercooled into the N-CDW phase. Inset: SEM image of device 3-a with 4-probe a.c. resistivity measurement configuration. (**B, C**) Reconstructed current density maps of the as-cooled supercooled N-CDW state (B) and slow-cooled C-CDW (C) state respectively ($I_{read}$ = 33.4 A/cm$^2$). The current spreads over the entire channel in both the states. (**D**) 4-probe a.c. transport tracked along with the application of d.c. pulses ($I_{write.}$) with increasing amplitudes. For a threshold $I_{write}$ = 16.7 x 10$^4$ A/cm$^2$ (10ms), the resistivity of the C-CDW state (red-orange) drops to ~ 60% of its as-cooled value (~18 mΩ-cm), pushing the system into the hidden state. For a maximum $I_{write}$ = 33.4 x 10$^4$ A/cm$^2$ (10 ms), the resistivity reaches ~ 50% of its as-cooled value. The resistivity in the supercooled N-CDW (blue-green curve) doesn't show any significant changes even for a higher pulse of $I_{write}$ = 40.1 x 10$^4$ A/cm$^2$ (10 ms). (**E, F**) Reconstructed current density maps of the supercooled N-CDW and slow-cooled C-CDW states after application of $I_{write}$, respectively. There is no significant change in the current flow in comparison to the as-cooled supercooled N-CDW state (E). The current in the pulsed C-CDW state exhibits behavior of the hidden state flowing along a confined channel (F).

It is unlikely that the DMCs are related to the FIB microstructuring. In Fig. 3A we show a flake of 1T-$TaS_2$ exfoliated onto a sapphire chip with pre-patterned contacts. As shown in Fig. 3B, on the first cooldown, the current profile looks homogeneous, just as in our FIB devices in Fig. 1D. On pulsing, a DMC is formed connecting the current contacts sticking to the edge of the flake, Fig. 3C, establishing that the DMCs appear in the C-CDW state of flake undisturbed by FIB processing (see also supplementary Fig. S7 G-I). Notably, the DMC does again form along a physical boundary as in the FIB devices. On changing the point of injection of the current, it can be seen that the DMC chooses the opposite physical boundary, Fig. 3D. (We note that while this behavior is by far the most common in all of our samples, we have observed in one of our devices a DMC form through the bulk of the flake, suggesting that physical edges are generally, but not exclusively, preferred, see supplementary Fig. S7 G-H). This suggests that the DMC will generally find natural boundaries to follow, whether in FIB devices or in flakes, and the choice of boundary depends on the nature of the charge injection.

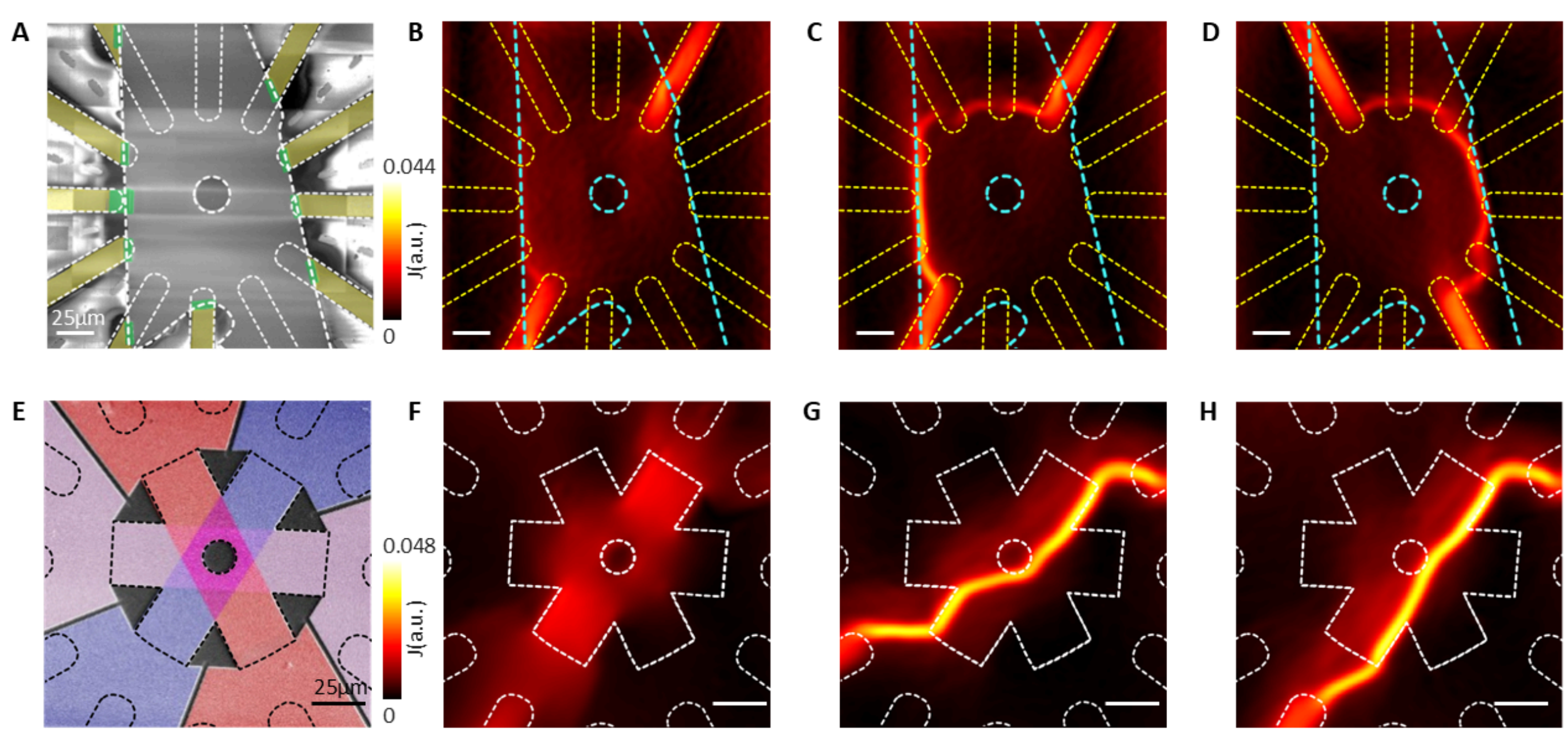

**Fig. 3. Role of flake edges, FIB and injection points to the DMCs configuration.** (**A**) SEM image of a 1T-$TaS_2$ device (device 4). (**B**) Reconstructed current density map in device 4 ($I_{read}$ = 5 μA), in the as-cooled C-CDW state. (**C, D**) Reconstructed current density maps after application of d.c. pulses ($I_{write}$ = 80mA, 10ms) along two different pairs of electrodes. The confined current in the hidden state, chooses to flow along different paths for the different electrode configurations. (**E**) SEM image of device 5. The false colors indicate the different pulse channels that are available. (**F**) Reconstructed current density map of the as-cooled C-CDW state in device 5 ($I_{read}$ = 8.3 A/cm$^2$). The current flows along the entire channel. (**G, H**) Reconstructed current density maps in device 5 after application of d.c. pulse ($I_{write}$ = 7.5 x $10^4$ A/cm$^2$,10 ms) between different electrodes. Along a defined channel in the device, the path carved by the confined current is affected by the choice of the pulse electrodes connected to the channel. The pulse current ($I_{write}$) and the probe current ($I_{read}$) were sent through the contacts that are visible in (B-D, F-H). Between images (C) and (D), (G) and (H), $I_{erase}$ (40 mA,100 ms for device 4; 3.33 x $10^4$ A/cm$^2$, 100 ms for device 5) was used to switch the devices 4 and 5 respectively, back to their high resistive C-CDW state, as described in Fig. 1.

The dependence on charge injection as well as the confined nature of the DMCs are reminiscent of the formation of the filaments created by Joule heating observed in other materials near a metal-insulator transition (*9*, *16–18*). Filamentary conduction paths in $VO_2$ and $V_2O_3$ have been observed when a current is passed, and understood as a cascade effect originating from local Joule heating (which has interesting percolation physics of its own (*19*)). However, in these cases the filaments disappear when the excitation is removed, in contrast to other non-volatile oxide switchable systems (*20*, *21*), and the DMCs we observe in 1T-$TaS_2$ (supplementary Fig. S5). The filaments in $VO_2$ also disappear at the MIT, whereas in 1T-$TaS_2$ the filaments are connected to the hidden state transition at $T_H$ (supplementary Fig. S6), and distinct from the C-CDW metal-insulator transition occurring at much higher temperatures. The DMCs are created by applying *higher* pulse currents in the high resistance state of the sample and erased by applying *lower* pulse currents in the low resistance state of the sample. When examining the threshold pulse width for writing-related processes, the Joule heating is significantly smaller than the erasing current (see supplementary Fig. S8 and Appendix A for estimated Joule heating). Combine this fact with the observation that the supercooled devices, which have similar resistance to the samples with DMCs, should exhibit a similar level of Joule heating yet show no evidence for the formation of DMCs (Fig. 2). We conclude that while Joule heating likely assists in the formation of the DMCs, it cannot be the sole driving mechanism. The origin of the DMC is directly connected to the formation of the hidden state.

Our data shows that the location of the DMCs will generally prefer boundaries, and depends sensitively on the location of the charge injection and interfaces present in the sample. This already points to a connection of the DMCs, and by association the hidden state, to defects present in the material. Such a picture has already been suggested by detailed STM measurements, which have recently shown that applied voltages can move and even annihilate pairs of topological defects in the C-CDW (*22*). We suggest that the DMCs are characterized by the alignment of such defects, creating conducting filaments in places where defects are favorable: boundaries (supplementary Fig. S7 (R-W). Since the defects come in non-polar pairs (*22*), this picture also answers why the location of the DMCs does not depend sensitively on polarity (see supplementary Fig. S9 for our complete investigation). Moreover, this picture allows a natural mechanism for metastability arising from a balance of the energy of the bound topological defects (which favors co-location at a boundary), and interactions between defects (which may favor a more uniform distribution across the system). The current pulses allow the manipulation of this balance through the electrical response of the C-CDW defects.

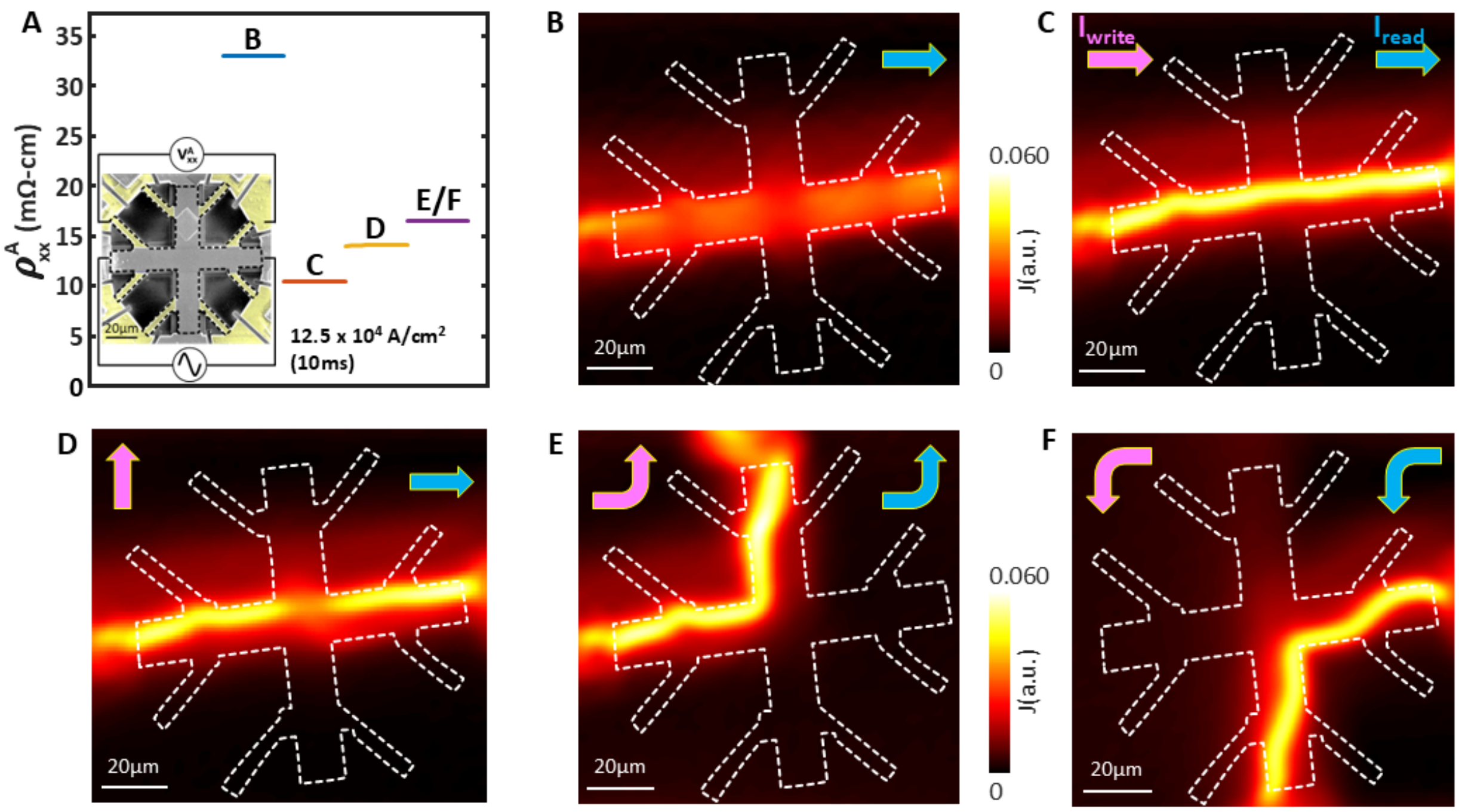

**Fig. 4. Design of a multi-resistance DMC switch.** (**A**) Resistivity values for different states in $1T\text{-}TaS_2$ (device 6), measured via a 4 probe a.c. resistivity configuration shown in the inset. The high resistivity state labeled B is the as-cooled C-CDW state. Resistivity states labeled C, D and E/F are created by applying pulses ($I_{write}$ = 12.5 x $10^4$ A/cm$^2$, 10 ms) along different pulse lead configurations, independently, from the as-cooled C-CDW state. The pulse current paths are indicated as pink arrows on the top-left corner of respective panels (C, D, E, F). Inset: SEM image of device 6. (**B**) Reconstructed current density map of the as-cooled CDW state in device 6. The current flows along the entire channel. The blue arrow on the top-right corner indicates the path of the alternating probe current ($I_{read}$ = 41.6 A/cm$^2$). (**C-F**) Reconstructed current density maps for the different pulsed metastable hidden states created by different pulsing ($I_{write}$ = 12.5 x $10^4$ A/cm$^2$, 10 ms) along different lead configurations while retaining the 4 probe a.c. resistivity configuration as shown in the inset of (A). The blue arrow on the top-right corner of each panel indicates the path of the alternating probe current ($I_{read}$ = 41.6 A/cm$^2$) for obtaining the current density maps. The confined current flows along different paths created by the pulsing channel definition. These metastable states are non-volatile (supplementary Fig. S5) and can be accessed by choosing the correct pulse lead configuration and toggling between the high resistivity and low resistivity states by the application of d.c. pulses. Between images (C), (D), (E), (F), the device was restored to its high resistive C-CDW state via application of $I_{erase}$ (6.25 x $10^4$ A/cm$^2$, 10 ms) pulses.

The microscopic mechanism of the dynamics of the DMCs notwithstanding, the knowledge gained in this experiment allows the simple design of devices and current protocols that can switch and orient the DMCs completely electrically. In the inset of Fig. 4A we show a bidirectional device whereby memory can in principle be encoded electrically as high and low resistance states. This is reminiscent of the bidirectional switching seen in antiferromagnetic spintronics designs (*23*, *24*), though the contrast between states is a factor of 2-4 in our system, whereas AFM systems show only a change of a few percent. In Fig. 4B-F we show current protocols that can run the DMCs along different edges, or even switch between physical boundaries, leading to multi-state resistance encoding as shown in Fig. 4A. This speaks to the significant flexibility of utilizing DMCs to create multi-state memory and communication protocols, leveraging the electrical control and non-volatility of the hidden state conductions paths. Metastable filamentary current paths have been proposed as “synaptic” connections in neuromorphic computing technologies (*18*). Here the DMCs of the hidden states demonstrate the ability to create and manipulate boundary states to encode a non-volatile synaptic network, where each piece of information corresponds to a unique DMC path running across the device.

The connection of the DMCs to the hidden state might help answer some long-standing questions in the physics of 1T-$TaS_2$ and the MIT. The absence of a heat capacity (or any thermodynamic) anomaly at $T_H$ is natural if the hidden state is associated with the mobility of topological domain defects. We can also make a stark distinction between the hidden state with the more conducting N-CDW state, which cannot host DMCs, consistent with the connection to defects arising only from domains of the C-CDW. It seems reasonable to expect that the ‘insulating’ C-CDW state has networks of more granular DMCs, rendering the distinction between metal and insulator a matter of the connectivity of the network, which is perhaps more distributed in the insulating state, but more of a ‘highway’ in the metallic state. In fact, recent theoretical approaches of novel metals emerging near a phase transition are reminiscent of this idea, emphasizing the importance of a network of spatially disordered interactions connecting conducting quasiparticles and an order parameter (*25*). This could account for the peculiar behavior of 1T-$TaS_2$, which includes surprising Kondo screening (*26*) that would be enabled by an ever-present network of disordered conduction channels, and the ‘cascade’ of metastable states (*10*), which could arise from variable network connectivity. Visualizing the connection between the DMCs running across and between boundaries, makes obvious how these states could be used in neural computing applications, leveraging the ability to electrically write a DMC of arbitrary complexity, but to read it with a simple resistive measurement.

***Methods:***

**Crystal synthesis:**

Single crystals of 1T-$TaS_2$ were grown via vapor transport using iodine as the transport agent. Stoichiometric amounts of tantalum and sulfur were mixed with 2.2 mg/$cm^3$ (growth A and B) or 5 mg/$cm^3$ (growth C and D) of iodine and sealed in a quartz ampoule, which was loaded into a horizontal two-zone furnace. The source end was held at 1050°C (growth A and B) or 950°C (growth C and D) and the sink end at 950°C (growth A and B) or 850°C (growth C and D) for a period of 7 days. High-quality crystals in dimensions excess of 1-2 mm were obtained after quenching the growth in ice water. Crystal structure and elemental stoichiometry were confirmed with powder X-ray diffraction and energy dispersive X-ray spectroscopy.
The C-CDW state can be reached by cooling at rates greater than 5 K/min in growths A, C and D. For growth B, the C-CDW state is reached only by cooling at a slow rate of 0.5 K/min. For a faster cooling rate, the device stays in the supercooled N-CDW state.

**Device fabrication:**

Focused Ion Beam (FIB) fabrication was performed by exfoliating single crystals of thicknesses under 10 μm and lateral dimensions of a few hundred micrometers onto sapphire substrates. 200 nm of gold was subsequently deposited over the sample and substrate. Using a Helios G4 FIB, gold was milled from the active area of the device, leaving a layer covering the sapphire substrate and edges of the crystal for electrical contacts. Platinum patches were added (via FIB deposition) connecting the edges of the device and the gold on the substrate (supplementary Fig. S2). The FIB was then used to mill the negative regions of the device and form the desired pattern. Contact resistances were of the order of a few ohms.

**Scanning SQUID measurements:**
Scanning SQUID microscopy was used to spatially map the current flow in the 1T-$TaS_2$ devices. We drive alternating probe currents ($I_{read}$ = 1-10 μA, 5 - 111 A/$cm^2$ for typical device dimensions, 0.9-2 kHz) between different leads of the sample to probe, and use the SQUID to map the magnetic field generated by the current flow. We reconstructed the current density using Biot-Savart's law to visualize the current distribution in the sample. The pulsing currents ($I_{write}$ and $I_{erase}$) density used for the switching in the measured devices, lies between $0.5x10^4$ - $44.4x10^4$ A/$cm^2$, about 3 orders of magnitude higher than the probe current density. The scanning images were taken at 4.2 K, unless otherwise specified.

***Acknowledgement:***
T. R. D. and B. K. thank Shalev Gur for assistance with the scanning SQUID measurements. **Funding:** Work by T. R. D. and B. K. was performed with support from the European Research Council grant no. ERC-866236, the Israeli Science Foundation grant no. ISF-228/22, the DIP KA 3970/1-1 and COST action SUPERQUMAP CA 21144. Work by J. G. A., J. E. M . was performed with support through a MURI project supported by the Air Force Office of Scientific Research (AFSOR) under grant number FA9550-22-1-0270. S. C. H. was supported by the U.S. Department of Energy, Office of Science, Basic Energy Sciences, Materials Sciences and Engineering Division under contract DEAC02-05-CH11231 within the Quantum Materials program (KC2202). Work by D. Y., M. S. and E. M. was performed with support from the European Research Council grant no. ERC-101117478, the Israeli Science Foundation grant no. ISF-885/23 and the PAZY foundation grant no. 412/23. **Author contributions:** B. K., E. M and J. G. A. designed the experiments, T. R. D. and B. K. performed the scanning SQUID measurements and analyzed the data. D. Y., S. C. H. and E. M. synthesized the samples. D. Y., S. C. H., M. S. and E. M. fabricated and characterized the devices. All the authors discussed the results and contributed to writing the manuscript. **Competing Interests:** The authors declare no competing financial interests. **Data and materials availability:** All data needed to evaluate the conclusions in the paper are present in the paper and/or supplementary materials.

Supplementary materials for

# Dark Metastable Conduction Channels near a Metal-Insulator Transition

## Fig. S1: Sample information

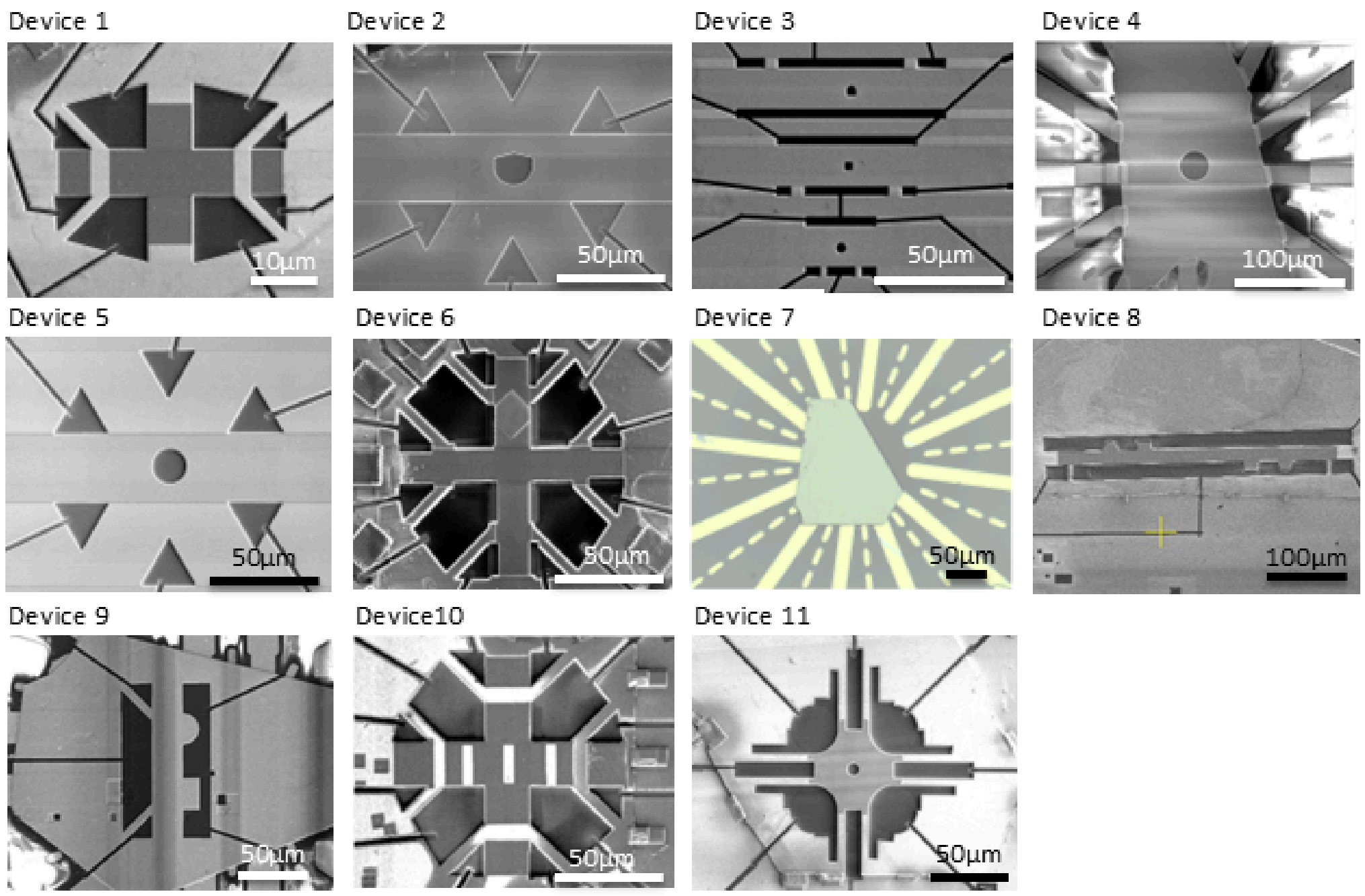

| Device number: | 1 | 2 | 3 | 4 | 5 | 6 | 7 | 8 | 9 | 10 | 11 |
|---|---|---|---|---|---|---|---|---|---|---|---|
| Growth: | A | A | B | A | C | A | A | A | A | A | D |
| Channel Width [µm] : | 20 | 30 | 15 | | 30 | 15 | | 16 | 20 | 20 | |
| Hole radius [µm] : | | 7.5 | | 26 | 15 | | | | | | 4 |
| Thickness [µm] : | 3.3 | 1.4 | 1.0 | 1.2 | 2.1 | 1.7 | 1.9 | 5.3 | 1 | 3.1 | 3.5 |

Scanning electron microscopy (SEM) images of the various devices (1 - 11) used in this study and a table containing the device dimensions. For details on growth, see the Methods section in the main text.

**Fig. S2: Device Fabrication**

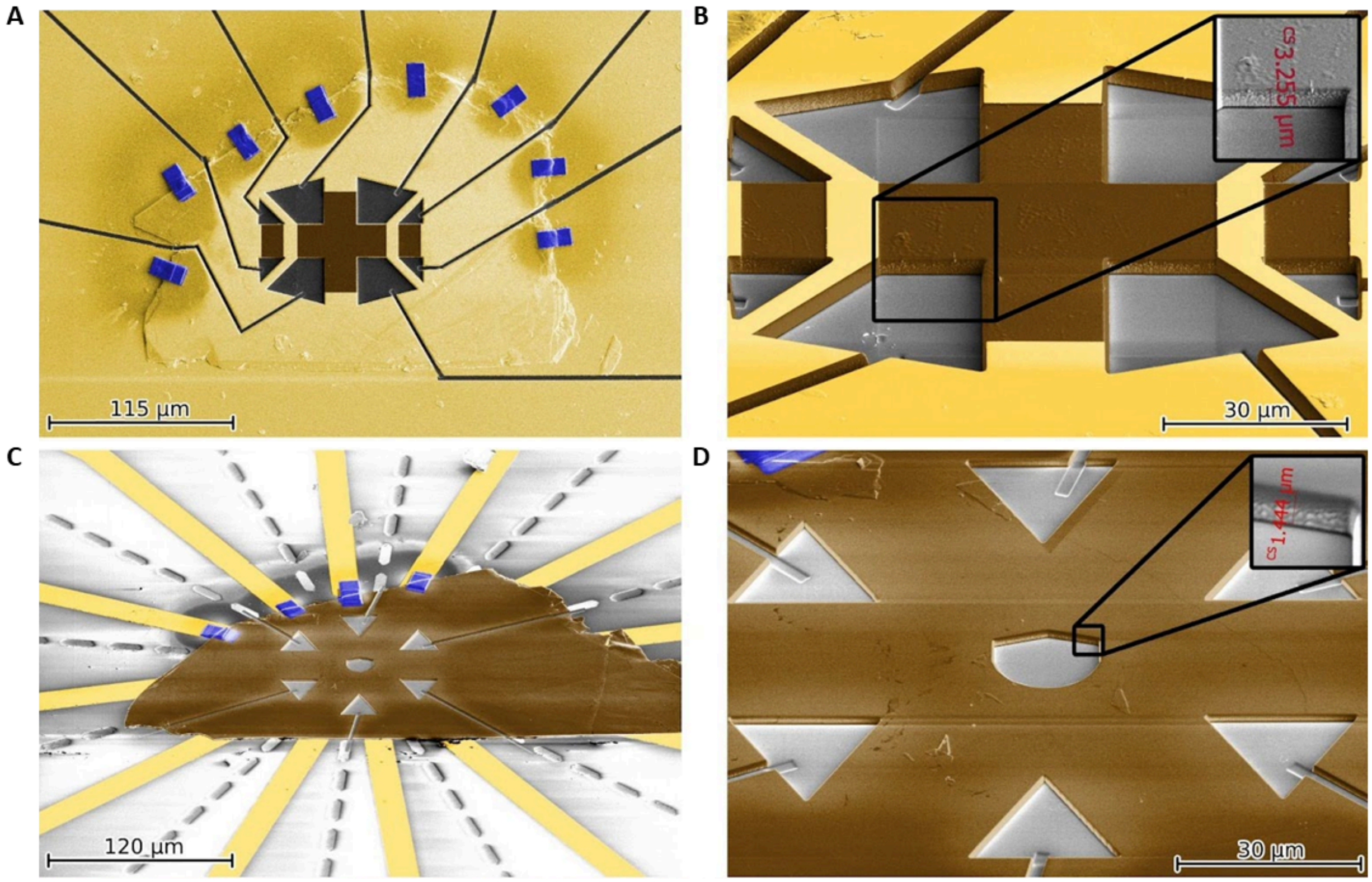

False color images of devices 1 and 2 where blue, copper, and brown represent platinum deposited by the FIB, gold contacts (or sputtered gold), and the 1T-$TaS_2$ crystal, respectively. The sapphire surface is not colored. (**A**) Device 1 with evaporated electrodes, where a cleaved sample is placed on top of a clear sapphire surface. Gold is sputtered over the sapphire. Using the FIB, the top layer of gold is removed from the device, then milled down to the sapphire surface to the desired pattern. Platinum contacts are deposited (using the FIB) on the edge of the crystal to improve the contacts of the gold. To create the contacts to the device, the sample is milled down to the sapphire surface which breaks the contact between different parts of the device, allowing the current to run only through the device (This milling is continued until the edge of the deposited gold). (**B**) Evaporated electrodes device at a higher magnification. The inset demonstrates the thickness measurements. (**C**) Device 2 with pre-patterned electrodes, where a cleaved crystal is placed on top of gold contacts on a sapphire substrate. The sample is milled using the FIB into the desired pattern until it reaches the sapphire surface. Platinum contacts are deposited using the FIB onto the gold contacts to improve (or connect) the contact with the sample. The dashed lines are used for alignment. (**D**) Pre-patterned electrodes device at a higher magnification, with an inset demonstrating how the thickness of the samples is measured. Thickness is measured on edges which are milled perpendicular to the surface.

**Fig. S3: Scanning SQUID raw data for Fig. 1 C-E.**

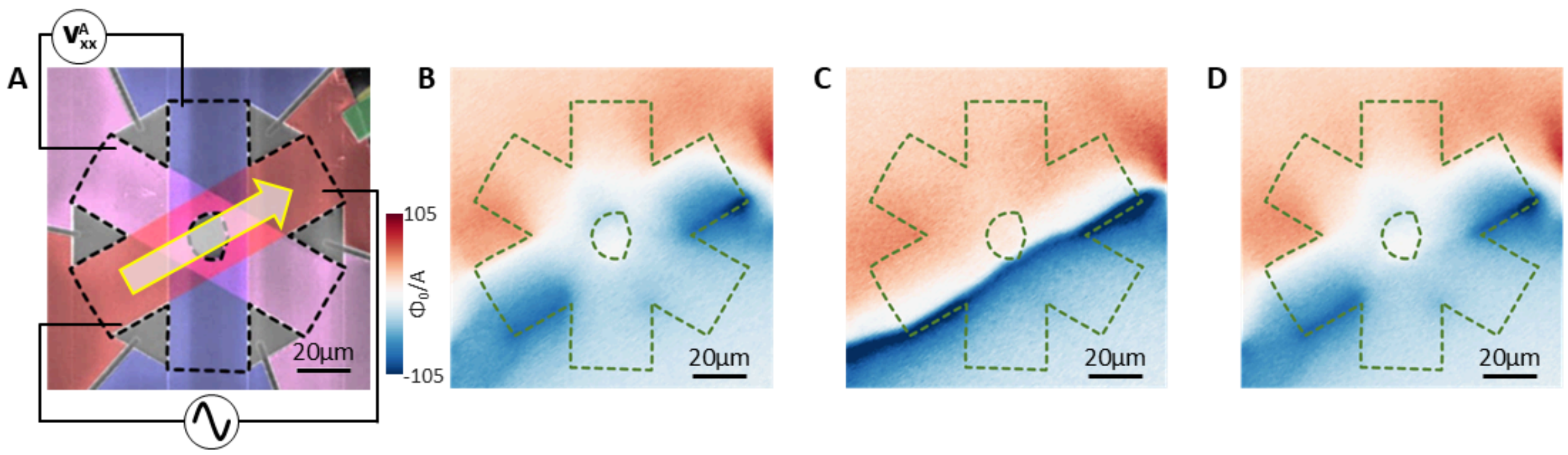

Images of the magnetic flux generated by the probe current ($I_{read}$) in device 2. (**A**) SEM image of device 2. False colors indicate the different channels in the device. Scanning SQUID image (raw data) of the magnetic flux generated by probe-current ($I_{read}$ = 12.61 A/cm$^2$) in the red channel in (**B**) as-cooled configuration at 4.2 K, (**C**) after application of write pulse ($I_{write}$ = 10.08 x 10$^4$ A/cm$^2$, 10ms) and (**D**) after application of erase pulse ($I_{erase}$ = 5.04 x 10$^4$ A/cm$^2$, 100ms). The arrow in (A) indicates the direction of the d.c. pulses.

**Fig. S4: Magnetic field distribution above current carrying sheets at different heights.**

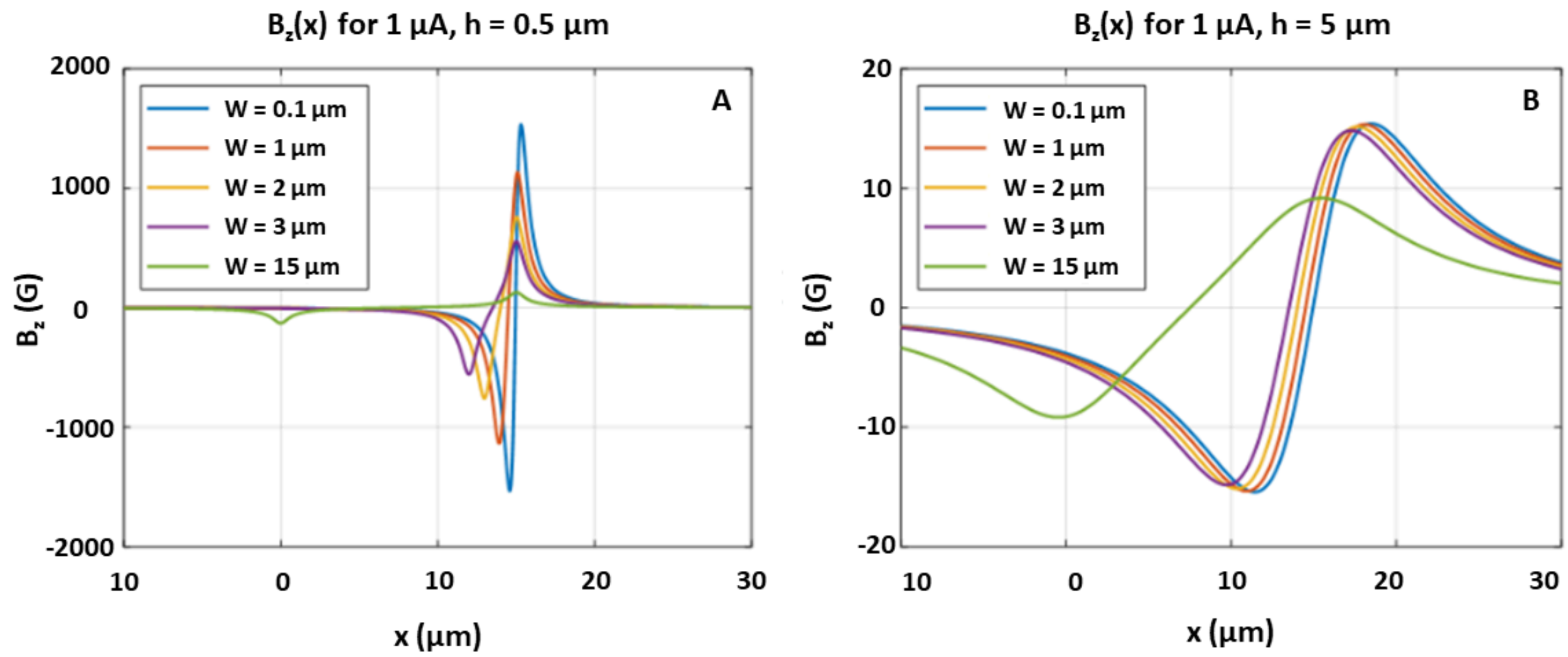

The plots illustrate the variation in magnetic field intensity $B_z(x)$, in Gauss as a function of the horizontal distance across the current-carrying infinite sheets. The magnetic field is calculated using the Biot-Savart law. We simulate the magnetic fields in infinite sheets of widths ranging from 0.1 µm to 15 µm, for 1 µA current. The magnetic fields are stronger and more focused for the narrower sheets. (**A**) $B_z(x)$ calculated for a sample-sensor separation of 0.5 µm. (**B**) $B_z(x)$ calculated for a sample-sensor separation of 5 µm.

**Fig. S5: Non-volatile nature of the conducting channels**

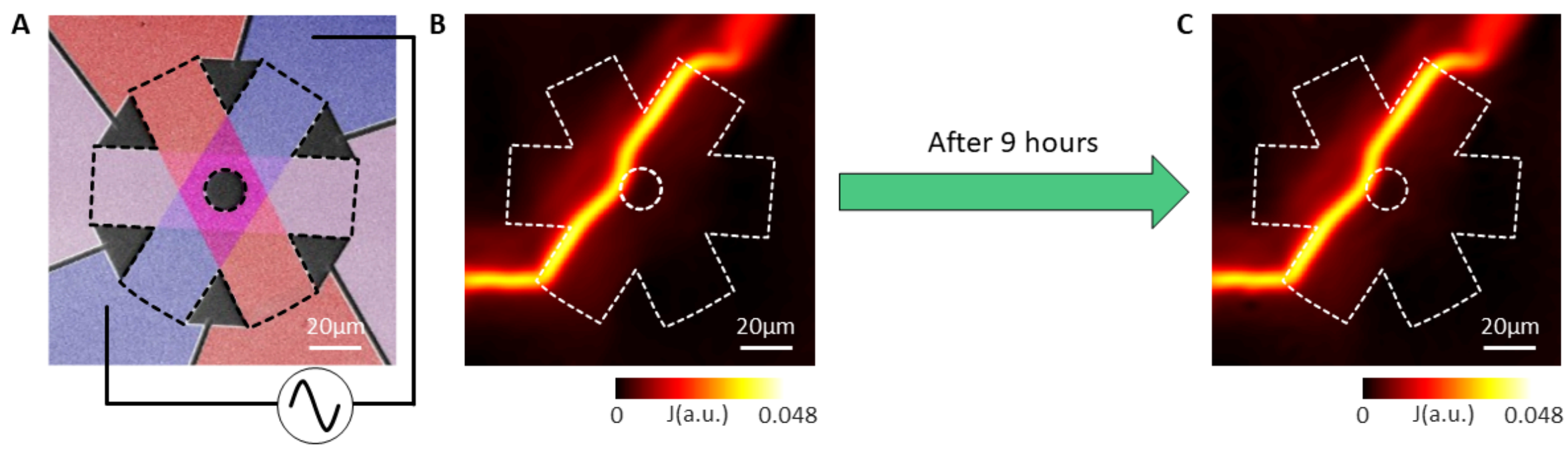

(**A**) SEM image of device 5. False colors indicate the different channels in the device. (**B**) Current density map of the conducting channel created by the application of a d.c. pulse ($I_{write}$ = 7.50 x $10^4$ A/cm$^2$, 10ms). (**C**) Current density map of the channel recorded after a time delay of 9 hours. The absence of any observable changes indicates that the conducting channels are stable for prolonged periods.

**Fig. S6: Erasure of DMC and no DMC formation for $T > T_H$**

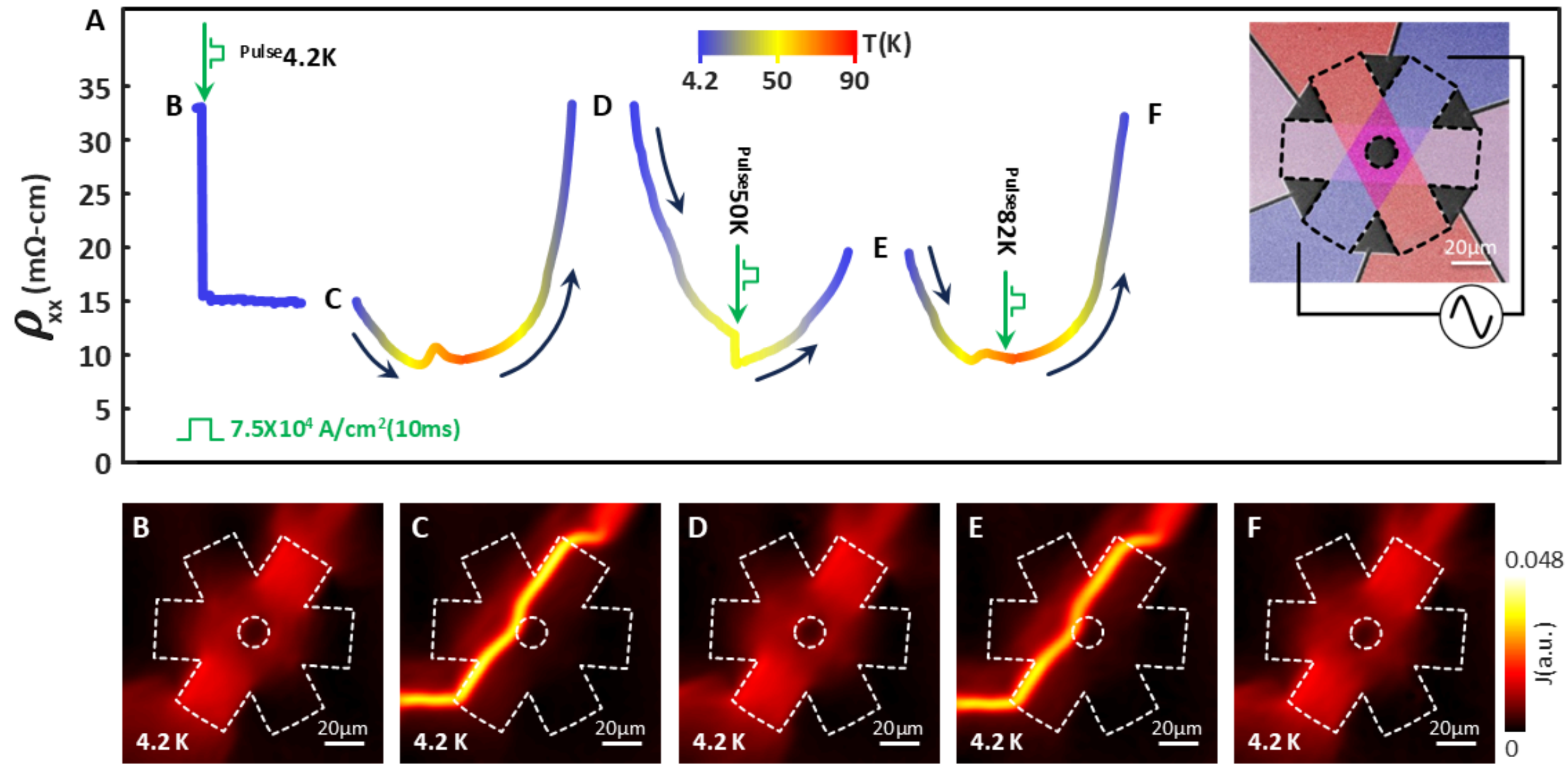

(**A**) Four-probe a.c. transport in device 5. The color gradient indicates the temperature of the device 5 as indicated by the temperature scale bar. The green square pulse and arrow indicate application of current pulses ($I_{write}$ = 7.5 x $10^4$ A/cm$^2$, 10ms). Inset: False color SEM image of device 5 with the $I_{read}$ source-drain leads indicated. The $I_{write}$ pulse leads are the same as the source-drain leads in the present measurement. (**B-F**) Reconstructed current density maps of the device at 4.2 K, obtained at different instances in the transport data, indicated in (A). (B) Current flows along the entire channel in the high resistivity C-CDW state. (C) Confined current channel in the low resistivity hidden state, after application of current pulse ($I_{write}$) at 4.2 K. (D) Recovery of the device into the high resistivity C-CDW state after cycling the device through $T > T_H$ ($T_H \sim$ 70 K). The current flows along the entire channel. (E) Confined current channel in the low resistivity hidden state, after application of current pulses ($I_{write}$) at 50 K. (F) Absence of confined current channel in the device at 4.2 K, after heating the device through $T > T_H$ to erase the DMC followed by application of current pulses ($I_{write}$) at 82 K.

**Fig. S7: Catalog**

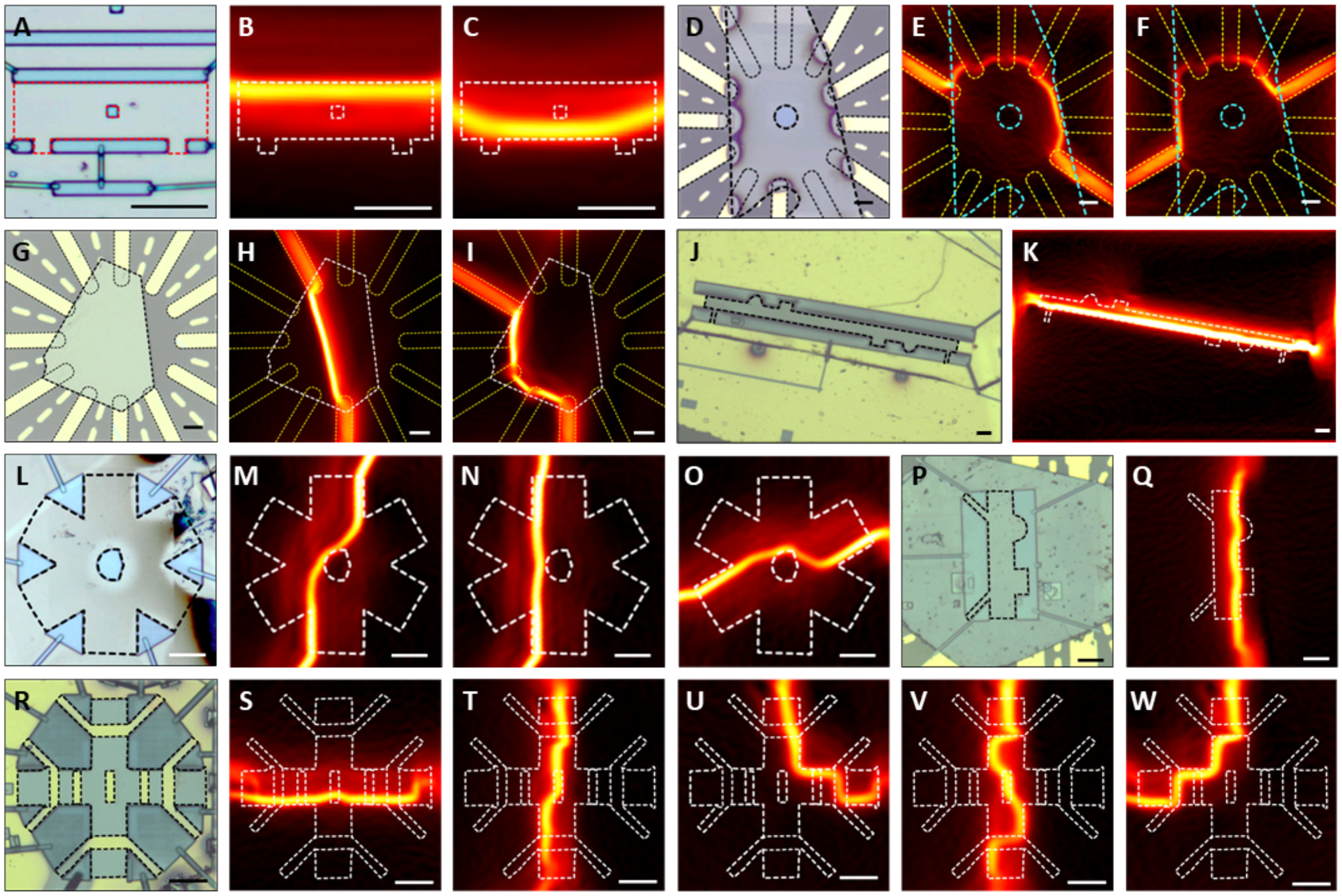

(**A**) Optical image of device 3-b. (**B**, **C**) Reconstructed current density maps of the conducting channel after application of pulses ($I_{write}$ = 33.4 x $10^4$ A/cm$^2$, 10ms). The conducting channels form on either boundary along the length of the channel by altering the initial choice of d.c. pulse leads. (**D**) Optical image of device 4. (**E**, **F**) Current density maps after application of $I_{write}$ pulses ($I_{write}$ = 80mA, 10ms). The channels flow either along the edge of the crystal or skip between gold leads. (**G**) Optical image of device 7. (**H**, **I**) Current density maps after application of $I_{write}$ pulses ($I_{write}$ = 55mA, 10ms). (H) is a single event of the conducting channel choosing the typical filamentary flow. (I) shows the channel flowing along the gold-1T-$TaS_2$ interface. (**J**) Optical image of device 8. (**K**) Current density map after application of pulses ($I_{write}$ = 4.12 x $10^4$ A/cm$^2$, 10ms). This device was fabricated with a long (~350μm) channel to investigate that the channel formation isn't a locally restricted phenomenon. (**L**) Optical image of device 2. (**M**-**O**) Current density maps after application of pulses ($I_{write}$ = 10.08 x $10^4$ A/cm$^2$, 10ms). (**P**) Optical image of device 9. (**Q**) Current density map after application of pulses ($I_{write}$ = 25 x $10^4$ A/cm$^2$, 10ms). (**R**) Optical image of device 10. (**S**-**W**) Current density maps after application of d.c. pulses ($I_{write}$ = 5.63 x $10^4$ A/cm$^2$, 10ms). When provided with multiple interfaces (gold-1T-$TaS_2$), the conducting channel chooses such crossings to traverse its length. Scale bar is 20μm.

**Fig. S8: Minimum pulse-width for creating and erasing DMC (device 6)**

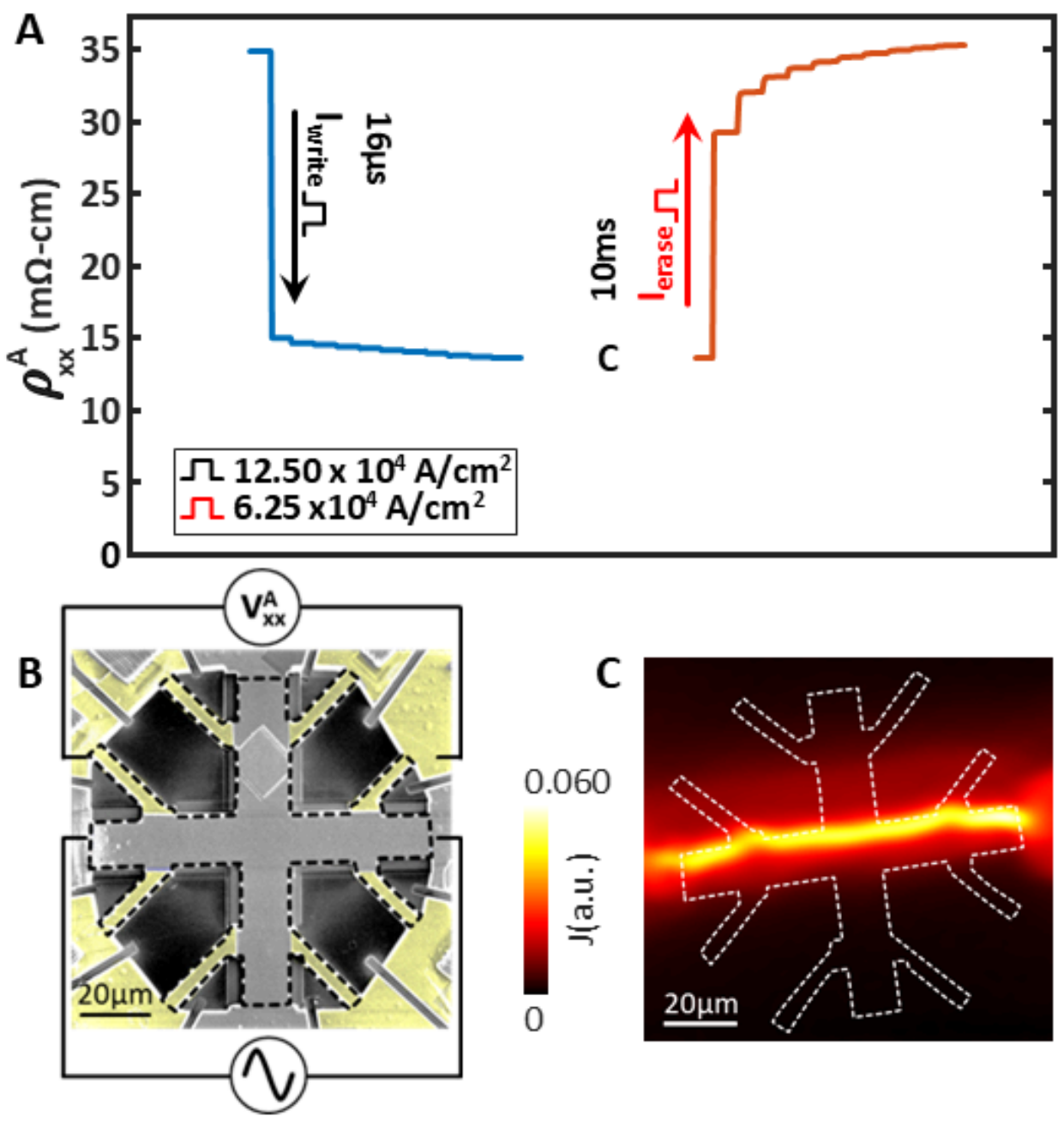

(**A**) Transport measurements of device 6 (4 probe). The pulse width is 16 μs for the “write” pulse and 10 ms for the “erase” pulse. The pulse amplitudes for “write” and “erase” are 12.50 x $10^4$ A/cm$^2$ and 6.25 x $10^4$ A/cm$^2$ respectively. (**B**) SEM image of device 6 with 4 probe measurement configuration. (**C**) Current density map of DMC created by “write” pulse

**Fig. S9: Effect of pulse polarity**

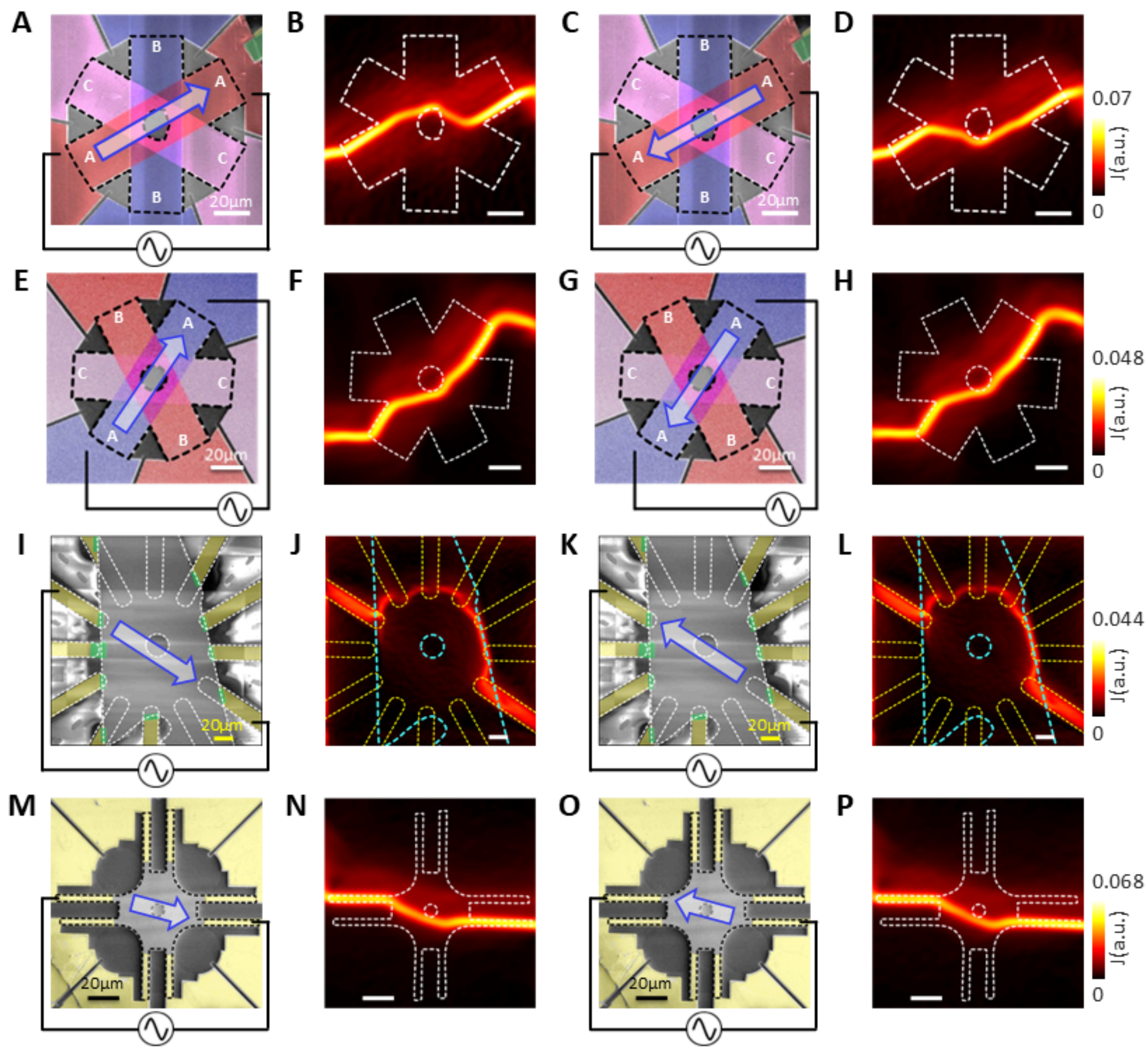

(**A, C**) SEM image of device 2 with false color overlays (blue, red and pink) identifying the different channels in the device. The blue-outlined white arrow indicates the direction of the $I_{write}$ pulse (10.08 x $10^4$ A/cm$^2$, 10ms). The pulse polarities are reversed between (A) and (C). **(B, D)** Current density maps of current after application of pulses. The DMCs choose different paths for the different polarities. (**E, G**) SEM image of device 5 with false color overlays (blue, red and pink) identifying the different channels in the device. The blue-outlined white arrow indicates the direction of the $I_{write}$ pulse (7.5 x $10^4$ A/cm$^2$,10 ms). The pulse polarities are reversed between (E) and (G). **(F, H)** Current density maps after application of pulses. There is no difference in the DMC paths for the different polarities. (**I, K**) SEM image of device 4. The blue-outlined white arrow indicates the direction of the $I_{write}$ pulse (80mA, 10ms). The pulse polarities are reversed between (I) and (K). **(J, L)** Current density maps after application of pulses. There is no difference in the DMC paths for the different polarities. (**M, O**) SEM image of device 11. The blue-outlined white arrow indicates the direction of the $I_{write}$ pulse (40mA, 10ms). The pulse polarities are reversed between (M) and (O). **(N, P)** Current density maps after application of pulses. There is no difference in the DMC paths for the different polarities. Scale bar is 20µm.

**Fig. S10: Anisotropy of channels, local view**

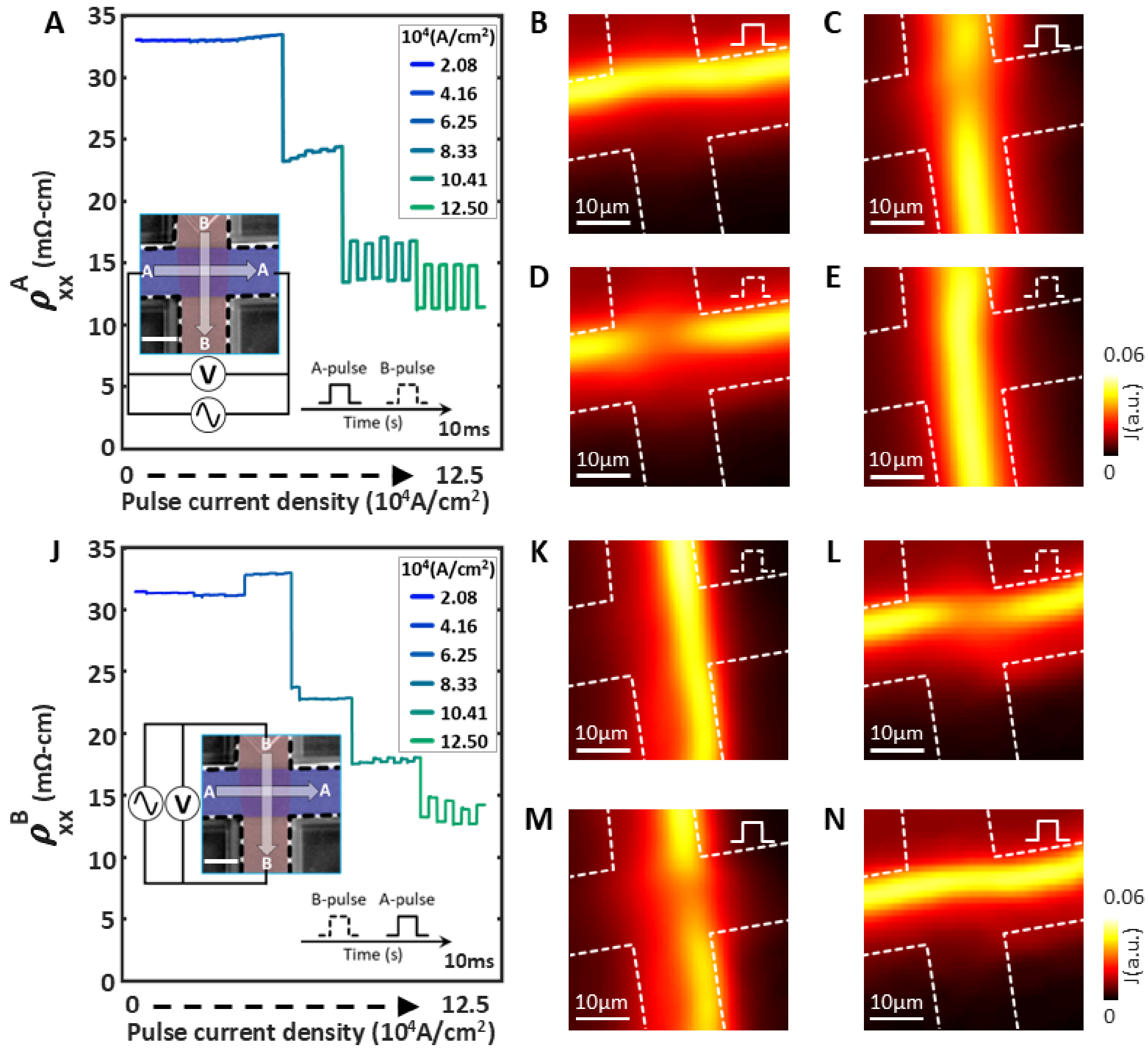

(**A**) Four-probe transport as a function of $I_{write}$ pulse density in cross-bar device 6. Inset: magnified view of the cross-bar section in the device and the a.c transport measurement geometry. The primary pulse direction is along A-bar. The d.c. pulses are applied alternatingly in A and B bars. The images are recorded after applying the strongest pulse ($I_{write}$ = 12.50 x $10^4$ A/cm$^2$, 10ms). (**B-E**) Current density maps show the channels at the cross-bar section of the device. (B, C) are images after pulsing along the A-bar and (D, E) after pulsing along the B-bar. (**J**) Four probe transport as a function of d.c. pulse density in cross-bar device 6. Inset - magnified view of the cross-bar section in the device and the a.c transport measurement geometry. The primary pulse direction is along B-bar. The d.c. pulses are applied alternatingly in B and A bars. The images are recorded after application of the highest pulse ($I_{write}$ = 12.50 x $10^4$ A/cm$^2$, 10ms). (**K-N**) Current density maps of the channels at the cross-bar section of the device. (K, L) were taken after pulsing along the B-bar and (M, N) after pulsing along the A-bar. The conducting channel in both of the above transport geometries is continuous at the cross-bar section in the direction of the applied d.c. pulse and is disrupted at the cross-bar section in the direction perpendicular to the applied d.c. pulse.

**Fig. S11: The current flows in the bulk of the sample**

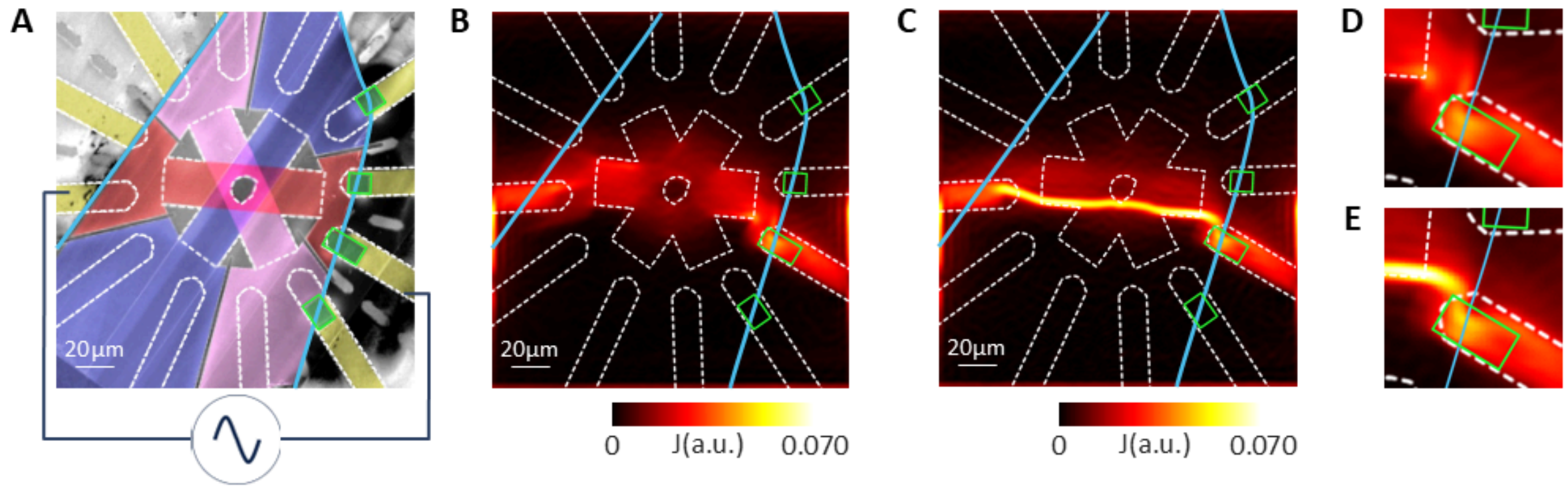

(**A**) SEM image of device 2 with false color overlays (blue, red and pink) identifying the different channels in the device. The blue outline traces the physical boundary of the crystal. The white dashed lines trace the pre-patterned gold electrodes (gold overlay) that contact the crystal from below. The green pads indicate the evaporated platinum staples deposited between the gold electrodes and the crystal. **(B-C)** Current density maps of current flowing in the channel with red overlay, in as-cooled (B) and after application of d.c. pulses ($I_{write}$ = 10.08 x $10^4$ A/cm$^2$, 10ms) (C). The conducting channel is observed flowing through the gold electrode/crystal interface across the length of the channel and into the platinum staples. **(D-E)** Magnified view of the device area containing the current carrying lead to the platinum staples in as-cooled (D) and pulsed (E) states

**Figure S12: No dependence on the probe current ($I_{read}$) direction**

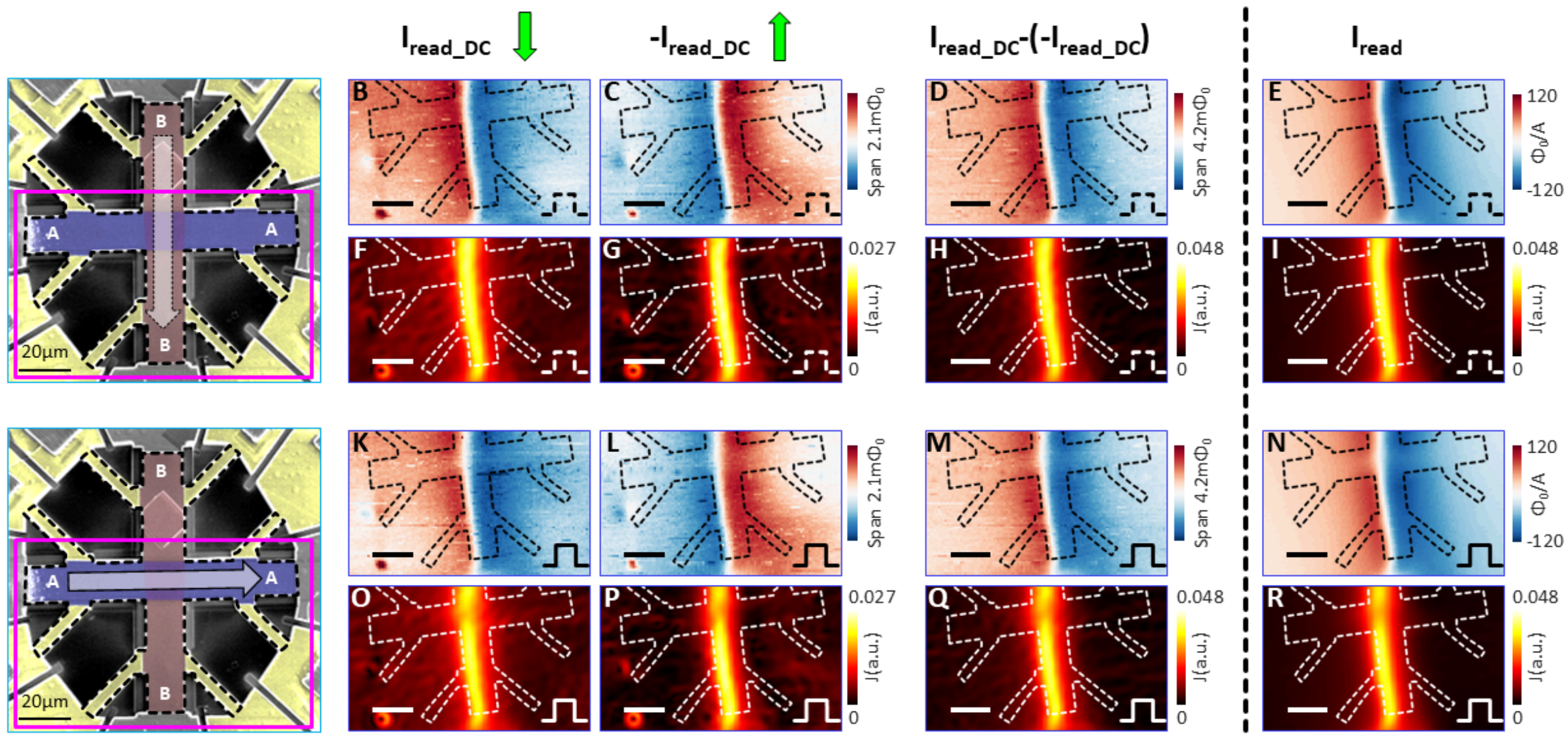

(**A**) SEM image of device 6 with false color overlays indicating the two pulse bars, A-bar (blue) and B-bar (red). After application of a d.c. pulse ($I_{write}$ = 12.5 x $10^4$ A/cm$^2$, 10 ms) along B-bar, the device is imaged using a probe current. Magnetic flux map of the region outlined by the magenta rectangle in (A) is recorded using a direct probe current ($I_{read_DC}$) of (**B**) 10 µA and (**C**) -10 µA. (**D**) Magnetic flux map created by subtracting (B) and (C). (**E**) Magnetic flux map recorded using an alternating probe current ($I_{read}$) of 10 µA. (**F-I**) Reconstructed current density maps of (B-E). (**J**) SEM image of device 6 with false color overlays indicating the two pulse bars, A-bar (blue) and B-bar (red). After application of a d.c. pulse ($I_{write}$ = 12.5 x $10^4$ A/cm$^2$, 10 ms) along A-bar,the device is imaged using probe current. Magnetic flux map of the region outlined by the magenta rectangle in (J) is recorded using direct probe current ($I_{read_DC}$) of (**K**) 10 µA and (**L**) -10 µA. (**M**) Magnetic flux map created by subtracting (K) and (L). (**N**) Magnetic flux map recorded using an alternating probe current ($I_{read}$) of 10 µA. (**O**-**R**) Reconstructed current density maps of (K-N). The direct probe current ($I_{read_DC}$) does not show any difference when applied in both directions. The differences, if any, would show up distinctly by decorating different paths when using an alternating probe current ($I_{read}$).

**Appendix A: Estimated heating generated during the pulsing process shown in this work.**

For each pulsing sequence, we estimate the Joule heating by $I^2Rt$; where I is the d.c. pulse amplitude in amperes; R is the sample resistance in ohms, just before the application of the first pulse of amplitude I; and t is the pulse width in seconds.

Few examples:

1. Device 6 - "write" threshold pulse and "erase" process in the C-CDW state (Supplementary Fig. S11).

Write pulse = $(0.030^2 \cdot 901 \cdot 0.000016) = 0.012$ mJ
Erase pulse = $(0.015^2 \cdot 353 \cdot 0.010) = 0.8$ mJ
The corresponding current density values for the pulses, in Fig. S11 are: 30 mA = $12.5 \times 10^4$ A/cm$^2$, 15 mA = $6.25 \times 10^4$ A/cm$^2$.

2. Device 3-a - "write" process in the C-CDW and N-CDW states (Fig. 2D).

C-CDW state = $(0.025^2 \cdot 600 \cdot 0.010) = 3.75$ mJ - resistivity switched from high to low, DMC generated.
N-CDW state = $(0.060^2 \cdot 182 \cdot 0.010) = 6.55$ mJ - nothing happened.
The corresponding current density values for the pulses, in Fig. 2D are: 25 mA = $16.7 \times 10^4$ A/cm$^2$, 60 mA = $40.1 \times 10^4$ A/cm$^2$.